\documentclass[aps,prl,twocolumn,superscriptaddress,showpacs,floatfix]{revtex4}

\usepackage{amssymb, amsmath, graphicx}

\begin{document}


\title{Evidence of Strong-Coupled Superconductivity in CaC$_6$ from Tunneling Spectroscopy}


\author{C.~Kurter}
\email[Corresponding author: ]{kurter@anl.gov}
\affiliation{Materials Science Division, Argonne National Laboratory, Argonne IL 60439}
\affiliation{Physics Division, Illinois Institute of Technology, Chicago, IL 60616}

\author{L.~Ozyuzer}
\affiliation{Materials Science Division, Argonne National Laboratory, Argonne IL 60439}
\affiliation{Department of Physics, Izmir Institute of Technology, Izmir, Turkey}

\author{Daniel Mazur}
\affiliation{Materials Science Division, Argonne National Laboratory, Argonne IL 60439}
\affiliation{Physics Division, Illinois Institute of Technology, Chicago, IL 60616}

\author{J.~F.~Zasadzinski}
\affiliation{Physics Division, Illinois Institute of Technology, Chicago, IL 60616}

\author{D.~Rosenmann}
\affiliation{Materials Science Division, Argonne National Laboratory, Argonne IL 60439}

\author{H.~Claus}
\affiliation{Materials Science Division, Argonne National Laboratory, Argonne IL 60439}

\author{D.~G.~Hinks}
\affiliation{Materials Science Division, Argonne National Laboratory, Argonne IL 60439}

\author{K.~E.~Gray}
\affiliation{Materials Science Division, Argonne National Laboratory, Argonne IL 60439}


\date{\today}

\begin{abstract}
Point-contact tunneling on CaC$_6$ crystals reproducibly reveals superconducting gaps, $\Delta$, of 2.3$\pm$0.2~meV which are $\sim$~40\% larger than earlier reports.  That puts CaC$_6$ into the class of very strong-coupled superconductors since 2$\Delta$/kT$_c\sim$~4.6.  Thus soft Ca phonons will be primarily involved in the superconductivity, a conclusion that explains the large Ca isotope effect found recently for CaC$_6$.  Consistency among superconductor-insulator-normal metal (SIN), SIS and Andreev reflection (SN) junctions reinforces the intrinsic nature of this result.
\end{abstract}


\maketitle


The discovery of superconductivity at 11.6~K in CaC$_6$ at ambient pressures\cite{Weller05} and up to 15.1~K at 8~GPa\cite{Gauzzi07} has reinvigorated interest in graphite intercalated compounds.  While the possibility of unconventional superconductivity had been suggested\cite{Csanyi05}, the collective experimental data on CaC$_6$ are consistent with weakly-coupled, electron-phonon driven superconductivity with a nearly isotropic energy gap. The recent scanning tunneling spectroscopy\cite{Bergeal06}, specific heat\cite{Kim06} and penetration depth measurements\cite{Lamura06} all indicate BCS weak-coupled superconductivity that is consistent with linear response theory\cite{Calandra05}.  However, Mazin et al.\cite{Mazin06} have pointed out several outstanding problems with the above picture including its inconsistency with recent data\cite{Hinks07} showing a large isotope effect for Ca and the less than perfect agreement with theory for the temperature dependence of the specific heat and the upper critical field, H$_{c2}$(T), data. The isotope effect and H$_{c2}$(T) dilemmas would disappear and the specific heat data might be better fit if CaC$_6$ were a strong-coupled superconductor.

Here we report evidence for strong coupling in CaC$_6$ based on point-contact tunneling (PCT) from both superconductor-insulator-normal metal (SIN) junctions and SIS junctions.  Both junction types reveal a gap parameter significantly larger than previously reported, yielding a strong coupling ratio, 2$\Delta$/kT$_c\sim$~4.6.  Andreev reflection spectroscopy (SN junctions) on the same crystals support this large ratio.  We will show that such a large strong coupling ratio, along with the modest T$_c$, indicates that soft phonon modes must dominate the pairing of electrons.  Low frequency modes in CaC$_6$ only involve Ca and, therefore, our results lead to a natural explanation of the large Ca isotope effect. This result may have broader implications by constraining the band structure and/or the calculation of gap anisotropy\cite{Sanna07}.

Preparation of CaC$_6$ used the alloy method as described by Emery et al.\cite{Emery05}.  A stainless steel (SS) tube is cleaned, baked at 900~$^\text{o}$C in vacuum and loaded with lithium and calcium in a 3:1 atomic ratio.  Natural, single-crystal graphite flakes are added and then the SS ampoule is mechanically sealed and placed inside a one-zone furnace, evacuated to 2$\times$10$^{-7}$~Torr, and then filled with argon gas. The reaction takes place in an argon atmosphere for 10 days at 350~$^\text{o}$C. The ampoule is then transferred to an argon glove bag where the alloy is remelted and the Ca intercalated graphite crystals are extracted.

\begin{figure}
\begin{center}
\includegraphics*[width=3.5in, bb=20 210 615 585]{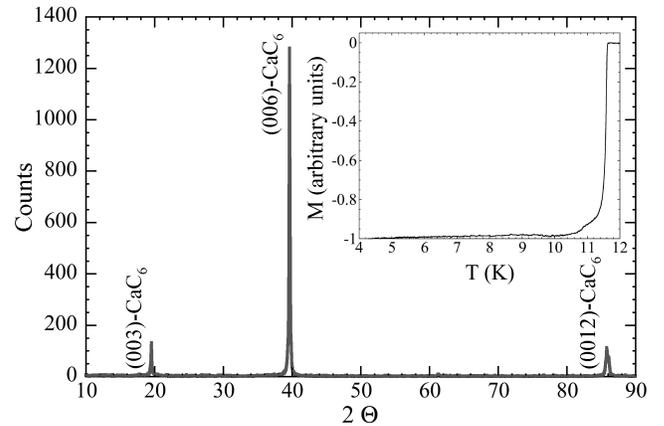}
\caption{\label{fig1}The (00$\ell$) diffraction peaks using Cu-K$_\alpha$ radiation. Inset: Normalized magnetization in a field of 0.1~G.}
\end{center}
\end{figure}

Figure~\ref{fig1} displays the x-ray diffraction pattern of a typical CaC$_6$ crystal showing the (00$\ell$) diffraction peaks obtained using Cu-$K_{\alpha}$ radiation taken in the Bragg-Brentano geometry.  No lines corresponding to hexagonal graphite are present within our detection limits confirming the bulk nature of the samples. Magnetization data, shown in the inset of Fig.~\ref{fig1}, were taken in a $\mu$-metal-shielded, non-commercial SQUID magnetometer on warming in a field of 0.1~G parallel to the $c$-axis of the sample after zero-field cooling.  They reveal a sharp, ultra-low-field superconducting transition onset at 11.6~K. 

Tunnel junctions were obtained at T=1.65~K using a PCT method described in detail elsewhere\cite{Ozyuzer98}.  The CaC$_6$ samples (and CaC$_6$ tips for SIS junctions) were mounted on a point contact tunneling system in a N$_2$ atmosphere and immediately cooled down to 4.2~K.  For SIN junctions, CaC$_6$ crystal flakes were oriented such that the vertical motion of the gold tip was nominally along the c-axis.  For SIS junctions a tip was fabricated from another CaC$_6$ crystal and aligned such that the tip and substrate crystals had similar $c$-axis orientations. Initial contacts of the tips in both configurations led to junctions of varying quality but with typical gap parameters, $\Delta$, smaller than 1.7~meV.  The relative ease of forming SIN and SIS tunnel junctions indicates that there is an insulating surface layer on CaC$_6$, which is presumably the high band-gap CaO.  After repeated contacts of tip and sample, likely resulting in some mechanical cleaning of the CaC$_6$ surface, junctions were found which had higher quality characteristics and gap parameters $>$\hspace{2pt}2~meV.  The proof of this low-temperature, effective\hspace{2pt}-ultra-high-vacuum cleaning of the surface was the fairly common appearance of Andreev reflection spectra (SN junctions) indicating a clean superconductor$-$Au junction.

\begin{figure}
\begin{center}
\includegraphics*[width=3.4in, bb=110 55 520 650]{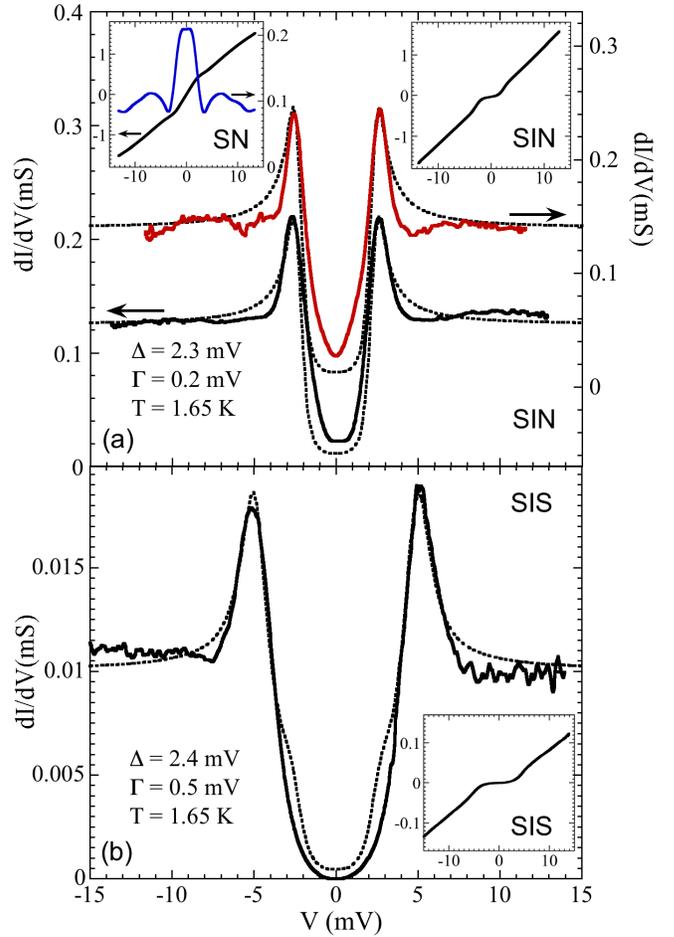}
\caption{\label{fig2}(Color online) (a) Two SIN dI/dV(V) are shown for different junctions at 1.65~K, each fit with an identical BCS curve. (b) One dI/dV(V) of an SIS junction at 1.65~K, along with the BCS fit.  Fits are using the parameters shown in respective plots.  Right-hand insets: Corresponding I(V) data, the I(V) curve in (a) corresponds to the lower dI/dV. Left-hand inset in (a): An example of measured Andreev reflection characteristics, I(V) and corresponding dI/dV(V) shown. All insets: Horizontal axis are voltages in mV, left vertical axis are currents in $\mu$A, right vertical axis, when present, is conductance in mS.}
\end{center}
\end{figure}

\begin{figure}
\begin{center}
\includegraphics*[width=3.5in, bb=15 120 615 690]{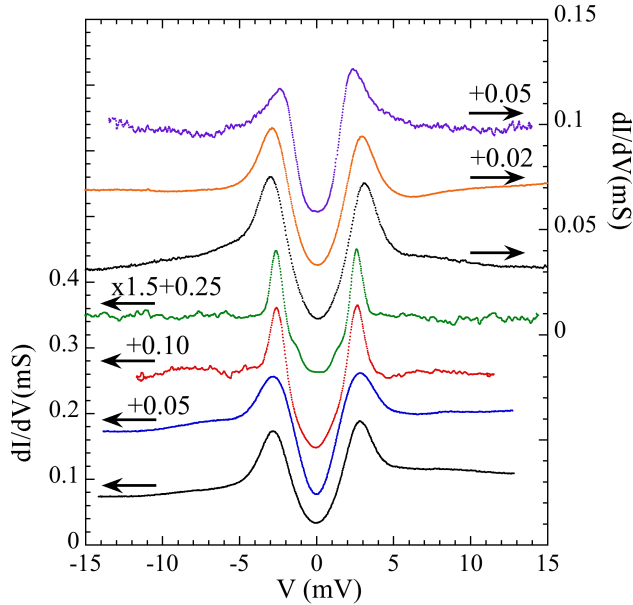}
\caption{\label{fig3}(Color online) A variety of SIN junctions made on three different crystals of CaC$_6$ showing consistent gap values of 2.1$-$2.5 meV.  }
\end{center}
\end{figure}

\begin{figure}
\begin{center}
\includegraphics*[width=3.35in, bb=18 77 662 705]{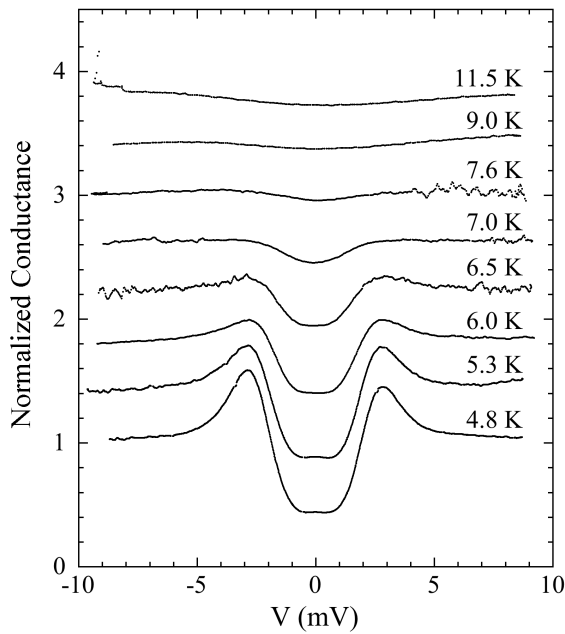}
\caption{\label{fig4}Temperature dependence of dI/dV is shown for one SIN junction.  }
\end{center}
\end{figure}

Figure~\ref{fig2} shows I-V and dI/dV vs. V for some of the high quality tunnel junctions (two SIN and one SIS) and one SN Andreev contact.  The well-defined conductance peaks for the SIS junctions are seen at twice the voltage of the SIN peaks as expected.  The large peak-height-to-high-voltage-background ratio as well as the relatively low zero-bias conductance (near zero for the SIS junction) attest to the high quality of the junctions.  Thus far we have never observed the flat sub-gap conductance expected for an isotropic $s$-wave energy gap, which may be an indication of gap anisotropy\cite{Sanna07}.  For convenience, fits of the raw conductance data to a BCS density-of-states modified (as in Ref.~\cite{Bergeal06}) with an empirical scattering rate\cite{footnote1}, $\Gamma$, are shown as dashed lines in Figs.~\ref{fig2}a and~\ref{fig2}b.  The gaps for these SIN and SIS junctions are 2.3 and 2.4~meV respectively.  The Andreev reflection data (Fig.~\ref{fig2}a, left inset) show the expected factor-of-two enhancement of the conductance below the energy-gap voltage for a clean metal-to-metal contact.  The SN energy gap can only be estimated to be 2.1$-$2.6 meV as there is no rigorous fitting procedure available for the entire spectrum.

Figure~\ref{fig3} displays a set of SIN tunneling data obtained on three different CaC$_6$ crystals.  As in previous reports\cite{Bergeal06}, the tunneling conductance shows no evidence of multiband superconductivity like the distinct energy gaps observed in MgB$_2$ (for a review see Ref.~\cite{Schmidt03}).  While the numerous junctions analyzed after initial surface cleaning exhibit a varying degree of quality, fits to all such SIN and SIS data lead to a range of gaps from 2.1 to 2.5~meV, with an average value of 2.3~meV.

We can easily rule out the possibility of extrinsic factors emulating the large gap we observe in this work.  For example, the SIN data could be from an inadvertent SIS break-junction caused by a piece of the CaC$_6$ crystal sticking to the Au tip.  In such a case, the peak at 2.3~mV would be the result of two pieces with an average $\Delta$=1.15~meV, a value much smaller than published range of 1.6$-$1.8~meV.  This is unlikely.  Furthermore the temperature dependencies of SIN and SIS junctions are very different, e.g., as T increases, SIS junctions reveal a closing of the 2$\Delta$ conductance peak, as well as the development of a conductance peak at zero bias due to thermally created quasiparticles.  The temperature dependence of one SIN junction with $\Delta$=2.3~meV is shown in Fig.~\ref{fig4}.  Although not as high a quality junction, the data show typical SIN behavior without any evidence for a zero-bias peak developing at higher T.

It is also possible that the SIS junctions, as in Fig.~\ref{fig2}b, are multiple junctions, e.g., two independent SIS junctions in series.  But in this case the observation of a single conductance peak would require that the two independent junctions have identical resistances, a highly unlikely scenario.  No such multiple-junction structure can be conceived for the Andreev reflection (SN) data.

Continuing this critical examination of the data, the increase in T$_c$ from 11.5~K to $\sim$15.1~K at a pressure of 8~GPa\cite{Gauzzi07} in CaC$_6$ forces us to consider the possibility that the larger $\Delta$ results from a larger T$_c$ brought about by local tip pressure.  Noting that the yield stress of high purity Au at 4.2~K is less than 100~MPa readily dispels this possibility\cite{footnote2}.  In addition, we find similar large gaps for SIS junctions without the soft Au tip.  Therefore we conclude that the common observation of $\Delta\sim$ 2.3~meV for SIN, SIS and SN junctions are not consequences of extrinsic factors.

The importance of this result is that it puts CaC$_6$ in the class of very strongly coupled superconductors (2$\Delta/$kT$_c\sim$~4.6), which implies that the electron-phonon coupling strength, $\lambda$, is significantly larger than the theoretical result\cite{Calandra05} of 0.83. Here $\lambda$ is an integral over $\alpha^2F(\omega)/\omega$, where $\alpha$ is the electron-phonon matrix element and $F(\omega)$ is the density of phonon states of frequency $\omega$. Clearly, $\lambda$ is heavily weighted by the low-frequency contributions, so the combination of large $\lambda$ and modest T$_c$ can only occur within standard strong-coupling theory of Allen and Dynes~\cite{Allen75} if their \mbox{T$_c$-equation} prefactor, $\omega_{log}$, defined therein, is very small.  It is useful to remember another strong-coupled superconductor, Pb, where the origins of the large strong coupling ratio and similar T$_c$ of 7.2~K are its low-energy (4$-$8~meV) phonons\cite{McMillan65}.  Using the empirical results on conventional superconductors\cite{Carbotte90}, a strong coupling ratio of 4.6 and T$_c$ of 11.6~K leads to a characteristic phonon energy, $\omega_{log}\sim$~5$-$7~meV.  In the case of CaC$_6$, such soft modes are a characteristic of the Ca intercalant only\cite{Calandra05,Mazin06}.  This would then suggest that Ca phonons are primarily involved in the pairing of electrons, a result which is quite different from the weak-coupling picture, but which would readily explain the nearly full isotope coefficient found with Ca isotopic substitution\cite{Hinks07}.

To date, the phonon density of states in CaC$_6$ has not been measured but phonon dispersions determined from band structure and linear response theory\cite{Calandra05,Mazin06} show in-plane zone boundary Ca modes near 10~meV.  It is possible that such modes could be renormalized to even lower frequencies due to strong electron-phonon coupling. Such renormalization is not included in linear response theory. The present work would argue that $\alpha^2F(\omega)$ is large at low frequencies and relatively small for higher phonon frequencies.  It would then imply that linear response theory\cite{Calandra05} is underestimating the strength of coupling to the low-frequency Ca modes.

Our proposed strong-coupled, low-frequency peak in $\alpha^2F(\omega)$ should be observable in the tunneling conductance, as for the classic case of Pb\cite{Allen75}.  We see a noticeable and reproducible negative deviation of the raw SIN data (Fig.~\ref{fig2}a) in high-quality junctions from the smeared BCS fit for $|eV|>\Delta$.  The characteristic energy expected from strong coupling to phonons is where this feature crosses over from a negative to positive deviation, and it is measured with respect to $\Delta$.  That energy is at most 5~meV, and although it would appropriately agree with the range of $\omega_{log}\sim$~5$-$7~meV estimated above, the present quality of tunneling data and our understanding of the normal state background are insufficient to confirm it or rule it out.

In summary, the superconducting gap of CaC$_6$ is found to be 2.3~meV.  This gap exceeds the previously reported value by approximately 40\%. Noting the consistency of SIN, SIS and SN junction configurations, and ruling out possible extrinsic causes, we conclude that this value is representative of the intrinsic energy gap of CaC$_6$.  The resulting strong coupling ratio of $\sim$4.6 changes considerably the current understanding of superconducting pairing in CaC$_6$. While this is a radical departure from the present consensus of weak coupling\cite{Bergeal06,Kim06,Lamura06,Calandra05}, it provides a direct explanation for the nearly full isotope effect found with Ca substitution\cite{Hinks07} and may help resolve the linear temperature dependance of H$_{c2}$(T).  This result may have a broader impact on band structure and electron-phonon calculations\cite{Mazin06}. Our result argues for intercalant-driven superconductivity with electron pairing dominated by the low-frequency Ca phonons; an idea that was proposed earlier\cite{Mazin05}.

\begin{acknowledgments}
This research was supported by the US Department of Energy, Basic Energy Sciences-Materials Sciences under Contract No. DE-AC02-06CH11357 at the Argonne National Laboratory operated by UChicago Argonne, LLC.
\end{acknowledgments}

\bibliographystyle{apsrev}

\begin{thebibliography}{20}
\bibitem{footnote2}
\bibinfo{note}{The room temperature value of the yield stress of high purity Au is about 40~MPa. Using the temperature dependence of the tensile stress\cite{McCammon57}, we estimate a yield stress of about 80~MPa at 4.2~K.  This is two-orders-of-magnitude lower than the 8~GPa needed to raise T$_c$ to 15.1~K.}
\end{thebibliography}

\end{document}